\documentclass[12pt,a4paper]{article}
\usepackage[utf8x]{inputenc}

\usepackage{amsmath}
\usepackage{amsfonts}
\usepackage{amssymb}

\usepackage{graphicx}

\usepackage[perpage]{footmisc}

\usepackage[bookmarks=true,colorlinks=true,citecolor=blue,linkcolor=blue,urlcolor=magenta]{hyperref}

\usepackage[numbers,sort&compress]{natbib}

\usepackage{multirow}

\usepackage[table]{xcolor}

\usepackage{geometry} 
\begin{document}

\author{V.D.~Rusov$^{1}$\footnote{Corresponding author: Vitaliy D. Rusov, E-mail: siiis@te.net.ua}, V.A.~Tarasov$^1$, V.M.~Vaschenko$^{2}$, E.P.~Linnik$^1$, \\ T.N.~Zelentsova$^1$, M.E.~Beglaryan$^1$, S.A.~Chernegenko$^1$, S.I.~Kosenko$^1$, \\ P.A.~Molchinikolov$^1$, V.P.~Smolyar$^1$, E.V.~Grechan$^1$}

\title{Fukushima plutonium effect and blow-up regimes in neutron-multiplying media}

\date{}

\maketitle

\begin{center}
  \textit{$^1$Department of Theoretical and Experimental Nuclear Physics, \\ Odessa  National Polytechnic University, Odessa, Ukraine}
  
  \textit{$^2$State Ecological Academy for Postgraduate Education, Kiev, Ukraine}
\end{center}

\abstract{It is shown that the capture and fission cross-sections of $^{238}$U and $^{239}$Pu increase with temperature within 1000-3000~K range, in contrast to those of $^{235}$U, that under certain conditions may lead to the so-called blow-up modes, stimulating the anomalous neutron flux and nuclear fuel temperature growth. Some features of the blow-up regimes in neutron-multiplying media are discussed.}


\section{Introduction}
\label{sec1}

It is known that after the loss of coolant at three nuclear reactors during Fukushima nuclear accident its nuclear fuel melted. It means that the temperature in the active zone at some moments reached the melting point of uranium-oxide fuel\footnote{Let us note that the third block partially used MOX-fuel enriched with plutonium.}, i.e. $\sim$3000$^{\circ}$C.

Surprisingly enough, scientific literature today contains absolutely no either experimental or even theoretically calculated data on behavior of the $^{238}U$ and $^{239}Pu$ capture and fission cross-sections depending on temperature at least in 1000-3000$^{\circ}$C range. At the same time there are serious reasons to believe that the cross-section values of the mentioned elements increase with temperature. At least we may point, for example, to the qualitative estimates by Ukraintsev~\cite{ref54}, Obninsk Institute of Atomic Energetics (Russia), that confirm the possibility for $^{239}Pu$ cross-sections growth in 300-1500$^{\circ}$C range.

Obviously, such anomalous temperature dependence of $^{238}U$ and $^{239}Pu$ capture and fission cross-sections may change the neutron and heat kinetics of nuclear reactors drastically. This is also true for the perspective new generation fast reactors (uranium-plutonium of Feoktistov~\cite{ref3} and thorium-uranium of Teller~\cite{ref4} type), that we classify as fast TWR reactors. Hence it is very important to know the anomalous temperature behavior of $^{238}U$ and $^{239}Pu$ capture and fission cross-sections, and furthermore it becomes critically important to know their influence on the heat transfer kinetics, since it may become a reason of the positive feedback\footnote{Positive Feedback is a type of feedback when a change in the output signal leads to the change in the input signal, which in its turn leads to a further deviation of the output signal from its original value. In other words, PF leads to instability and appearance of qualitatively new (often self-oscillation) systems.} (PF) with neutron kinetics leading to undesirable solution stability loss (the nuclear burning wave) as well as to a trivial reactor runaway with a subsequent nontrivial catastrophe.

A special case of PF is a non-linear PF, which leads to the system evolution in so-called blow-up mode \cite{ref55,ref56,ref57,ref58,ref59,ref60}, or in other words, in such a dynamic mode when one or several modeled values (e.g. temperature and neutron flux) grows to infinity at a finite time. In reality, a phase transition is observed instead of the infinite values in this case, and this can in its turn become a first stage or a precursor of the future technogenic disaster.

Investigation of the temperature dependence of $^{238}U$ and $^{239}Pu$ capture and fission cross-sections in 300-3000$^{\circ}$C range, and correspondingly, the heat transfer kinetics and its influence on neutron kinetics in TWR, is the main goal of the present paper.

\section{Temperature blow-up regimes in neutron-multiplying media}
\label{sec2}

Heat transfer equation for uranium-plutonium fissile medium is:

\begin{align}
\rho \left( \vec{r}, T, t \right) \cdot & c \left( \vec{r}, T, t \right) \cdot \dfrac{\partial T \left( \vec{r},t \right)}{\partial t} = \nonumber \\
& = \aleph \left( \vec{r},T,t \right) \cdot \Delta T \left( \vec{r},t \right) + \nabla \aleph \left( \vec{r},T,t \right) \cdot \nabla T \left( r,t \right) + q_T ^f \left( \vec{r},T,t \right),
\label{eq69}
\end{align}

\noindent where the effective medium density is

\begin{equation}
\rho \left( \vec{r},T,t \right) = \sum \limits_i N_i  \left( \vec{r},T,t \right) \cdot \rho_i,
\label{eq70}
\end{equation}

\noindent $\rho_i$ are tabulated values, $N_i \left( \vec{r},T,t \right)$ are the concentrations of the medium components, while the effective specific heat capacity (accounting for the medium components heat capacity values $c_i$) and fissile material heat conductivity coefficient (accounting for the medium components heat conductivity coefficients $\aleph_i (T)$) respectively are:

\begin{equation}
c \left( \vec{r},T,t \right) = \sum \limits_i c_i (T) N_i \left( \vec{r},T,t \right),
\label{eq71}
\end{equation}

\begin{equation}
\aleph \left( \vec{r},T,t \right) = \sum \limits_i \aleph_i (T) N_i \left( \vec{r},T,t \right).
\label{eq72}
\end{equation}

Here $q_T ^f \left( \vec{r},T,t \right)$ is the heat source density produced by nuclear fissions $N_i$ of fissile metal components which vary in time.

Theoretical temperature dependence of heat capacity $c(T)$ for metals is known: at low temperatures $c(t) \sim T^3$, and at high temperatures $c(T) \rightarrow const$, where the constant value ($const \approx 6 ~Cal/(mol \cdot deg)$) is determined by Dulong-Petit law. At the same time it is known that thermal expansion coefficient is small for metals, therefore the specific heat capacity at constant volume $c_v$ almost equals to the specific heat capacity at constant pressure $c_p$. On the other hand, the theoretical dependence of heat conductivity $\aleph_i (T)$ at high temperature of "fissile" metals is not known, while it is experimentally determined that the heat conductivity  coefficient $\aleph (T)$ of fissile medium is a non-linear function of temperature (e.g. see~\cite{ref61}, where heat conductivity coefficient is given for $\alpha$-uranium 238 and for metallic plutonium 239, and also~\cite{ref62}).

Further for solving thermal conductivity equation we used the following initial and boundary conditions:

\begin{equation}
T (r,t = 0) = 300~K ~~~and ~~ j_n = \aleph \left[ T(r \in \Re,t) - T_0 \right],
\label{eq73}
\end{equation}

\noindent where $j_n$ is the normal (to the fissile medium boundary) heat flux density component, $\aleph (T$) is the thermal conductivity coefficient, $\Re$ is the fissile medium boundary, $T_0$ is the temperature of the medium adjacent to the active zone.

Obviously, if the cross-sections of some fissile nuclides increase with temperature, then due to exothermic nature of the nuclei fission reaction, the significantly non-linear kinetics of mother and daughter nuclides in the nuclear reactor will immediately result in an autocatalytic increase of generated heat, just like in the autocatalytic processes of exothermic chemical reactions. The heat source density $q_T ^f \left(\vec{r},\Phi,T,t \right)$ which characterizes the amount of generated heat in this case will be:

\begin{equation}
q_T ^f \left(\vec{r},\Phi,T,t \right) = \Phi \left(\vec{r},T,t \right) \sum \limits_i Q_i^f \overline{\sigma}_f^i \left(\vec{r},T,t \right) N_i \left(\vec{r},T,t \right), ~~[W/cm^3],
\label{eq74}
\end{equation}

\noindent where

\begin{equation}
\Phi \left(\vec{r},T,t \right) = \int \limits_0 ^{E^{max}_n} \Phi \left(\vec{r},E,T,t \right) dE \nonumber
\end{equation}

\noindent is the total neutron flux density; $\Phi \left(\vec{r},E,T,t \right)$ is the density of a neutron flux with energy $E$; $Q_i ^f$ is the mean heat released in one fission event of the $i$-th nuclide;

\begin{equation}
 \overline{\sigma}_f^i \left(\vec{r},T,t \right) = \int \limits_0 ^{E_n^{max}} \sigma_f^i (E,T) \rho \left(\vec{r},E,T,t \right) dE  \nonumber
\end{equation}

\noindent is the  fission cross-section of the $i$-th nuclide averaged over the neutron spectrum;

\begin{equation}
 \rho \left(\vec{r},E,T,t \right) = \Phi \left(\vec{r},E,T,t \right) / \Phi \left(\vec{r},T,t \right)  \nonumber
\end{equation}

\noindent is the probability density function of the neutron energy distribution; $\sigma_f^i (E,T)$ is the microscopic fission cross-section  of the i-th nuclide, which is known to depend on the neutron energy and fissile medium temperature (Doppler effect~\cite{ref35}); $N_i \left(\vec{r},T,t \right)$ is the density of the $i$-th nuclide nuclei.

As follows from~(\ref{eq74}), in order to build a density function of the heat source, it is necessary to solve a problem related to the construction of a theoretical dependence of the cross-sections $\overline{\sigma}_f^i \left(\vec{r},T,t \right)$ averaged over the neutron spectrum on the temperature of reactor fuel (fissile medium). As is known, the impact of the nuclei thermal motion in the medium comes to a broadening and lowering of the resonances. By optical analogy this phenomenon is usually referred to as Doppler effect~\cite{ref35}. Since the resonance levels in the low energy region are observed for heavy nuclei only, the Doppler effect is noticeable only during the interaction of neutrons with such nuclei. And the higher is the environment temperature, the stronger is the effect.

\begin{figure}
  \begin{center}
    \includegraphics[width=10cm]{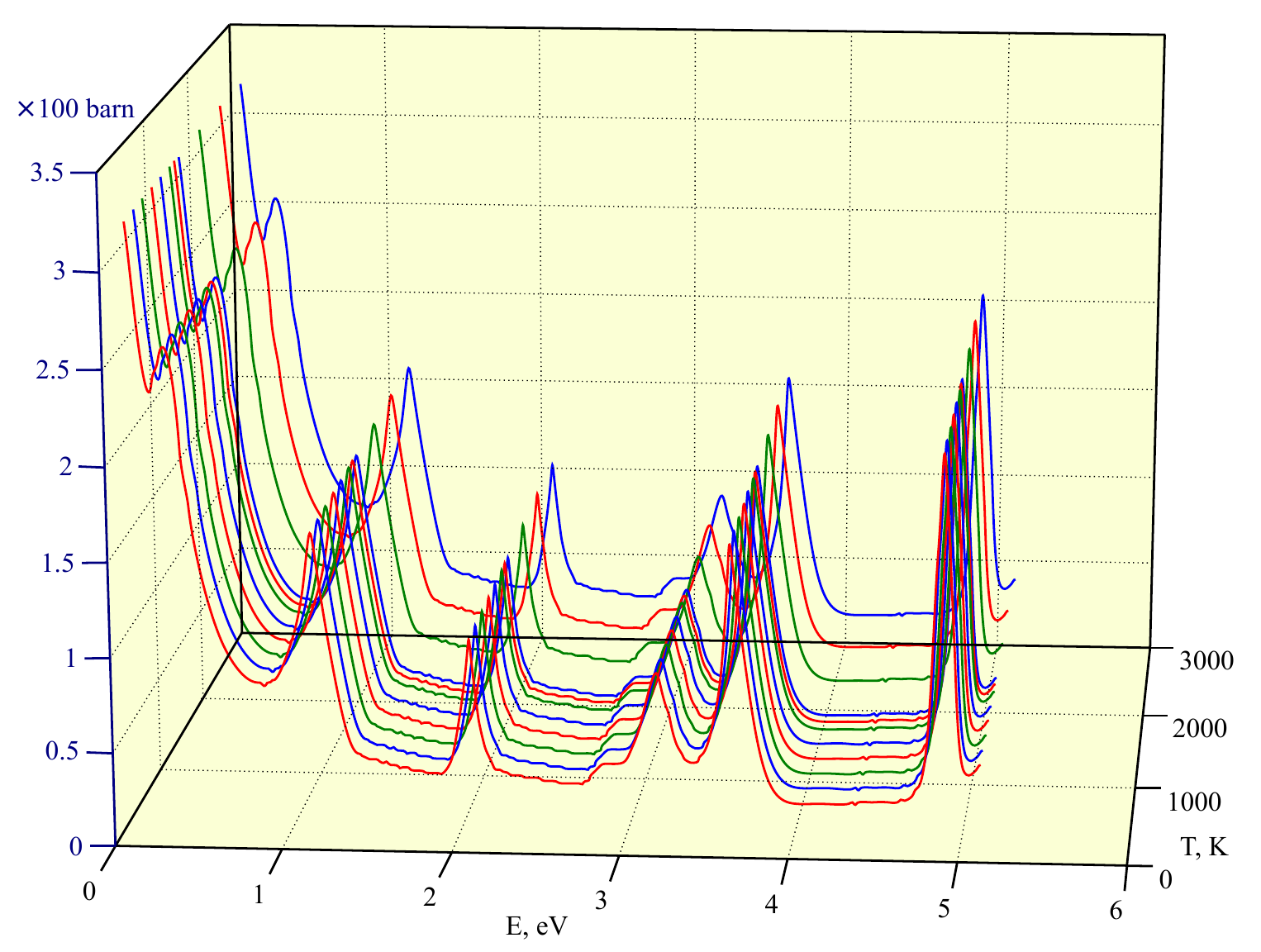}
    \caption{Calculated dependency of radioactive neutron capture reaction on  its energy for $^{235}_{92}U$ at different temperatures within 300K to 3000K.}
    \label{fig11}
  \end{center}
\end{figure}

Therefore a program was developed using Microsoft Fortran Power Station 4.0 (MFPS 4.0) that allows at the \underline{\textbf{first stage}} to calculate the cross-sections of the resonance neutron reactions depending on neutron energy taking into account the Doppler effect. The cross-sections dependence on neutron energy for reactor nuclides from ENDF/B-VII database~\cite{ref63} corresponding to 300K environment temperature were taken as the input data for the calculations. For example, the results for radioactive neutron capture cross-sections dependence on neutron energy for $^{235}$U are given in Fig.~\ref{fig11} for different temperatures of the fissile medium in 300K-3000K temperature range. Using this program, the dependence of scattering, fission and radioactive neutron capture cross-sections for essential reactor fuel nuclides $_{92}^{235}$U, $_{92}^{238}$U, $_{92}^{239}$U and $_{94}^{239}$Pu were obtained for different temperatures in 300K-3000K range.

At the \underline{\textbf{second stage}} a program was developed to calculate the dependence of the cross-sections $\overline{\sigma}_f^i \left(\vec{r},T,t \right)$ averaged over the neutron spectrum for main reactor nuclides and for main reactor nuclear neutron reactions for the specified temperatures. The averaging of neutron cross-sections for the Maxwell distribution was performed using the following expression:

\begin{equation}
\left \langle \sigma \left( E_{lim},T \right) \right \rangle = \dfrac{\int \limits_0 ^{E_{lim}} E^{1/2} e^{-E/kT} \sigma (E,T) dE}{\int \limits_0 ^{E_{lim}} E^{1/2} e^{-E/kT} dE}, \nonumber
\end{equation}

\noindent where $E_{lim}$ is the the upper limit of neutrons thermalization, while for the procedure of neutron cross-sections averaging over Fermi spectrum the following expression was used: 

\begin{equation}
\left \langle \sigma \left( E_{lim},T \right) \right \rangle = \dfrac{\int \limits_{E_{lim}}^{\infty} \sigma (E,T) E^{-1} dE}{\int \limits_{E_{lim}}^{\infty} E^{-1} dE}, \nonumber
\end{equation}

During further calculations in our programs we used the results obtained at the first stage i.e. the dependence of reaction cross-sections on neutron energy and medium temperature (Doppler effect). The neutron spectrum was specified in a combined way -- below the limit of thermalization $E_{lim}$ the neutron spectrum was described by Maxwell spectrum $\Phi_M \left( E_n \right)$; above $E_{lim}$ but below $E_F$ (upper limit for Fermi neutron energy spectrum) the neutron spectrum was described by Fermi spectrum $\Phi_F (E)$ for a moderating medium with absorption; above $E_F$, but below maximal neutron energy $E_n^{max}$  the spectrum was described by $^{239}Pu$ fission spectrum \cite{ref21,ref22}. Here the neutron gas temperature for Maxwell distribution was given by (\ref{eq75}), described in \cite{ref35}. According to this approach~\cite{ref35}, the drawbacks of standard slowing-down theory for thermalization area may be formally reduced if a variable $\xi (x) = \xi (1 - 2/z)$ is introduced instead of the average logarithmic energy loss $\xi$, which is almost independent of neutron energy (as is known, for environment consisting of nuclei with $A > 10$ the statement $\xi \approx 2/A$ is true). Here $z = E_n / kT$, $E_n$ is the neutron energy, $T$ is the environment temperature. Then within such a framework the following expression may be used for the temperature of the neutron gas in Maxwell spectrum of thermal neutrons\footnote{A very interesting expression revealing a hidden connection between the temperature of the neutron gas and the medium (fuel) temperature.}:

\begin{equation}
T_n = T_0 \left[ 1 + \eta \cdot \dfrac{\Sigma_a (k T_0)}{\langle \xi \rangle \Sigma_S} \right],
\label{eq75}
\end{equation}

\noindent where $T_0$ is the fuel medium temperature, $\Sigma_a (k T_0)$ is the absorption cross-section for energy $k T_0$, $\eta = 1.8$ is the dimensionless constant, $\langle \xi \rangle$ is averaged over the whole energy interval of Maxwell spectrum $\xi (z)$ at $kT = 1~eV$.

Fermi neutron spectrum for a moderating medium with absorption (we considered carbon as a moderator and $^{238}$U, $^{239}$U and $^{239}$Pu as absorbers) was set in the form~\cite{ref35,ref64}:

\begin{equation}
\Phi_{Fermi} \left(E, E_F \right) = \dfrac{S}{\langle \xi \rangle \Sigma_t E} \exp {\left[ - \int \limits_{E_{lim}}^{E_f} \dfrac{\Sigma_a \left( E' \right) dE'}{\langle \xi \rangle \Sigma_t \left(E' \right) E'} \right]},
\label{eq76}
\end{equation}

\noindent where $S$ is the total volume neutron generation rate, $\langle \xi \rangle = \sum \limits_i \left( \xi_i \Sigma_S^i \right) / \Sigma_S$, $\xi_i$ is the average logarithmic decrement of energy loss, $\Sigma_S^i$ is the macroscopic scattering cross-section of the $i$-th nuclide, $\Sigma_t = \sum \limits_i \Sigma_S^i + \Sigma_a^i$ is the total macroscopic cross-section of the fissile material, $\Sigma_S = \sum \limits_i \Sigma_S^i$ is the total macroscopic scattering cross-section of the fissile material, $\Sigma_a$ is the macroscopic absorption cross-section, $E_F$ is the upper neutron energy for Fermi spectrum.

The upper limit of neutron thermalization $E_{lim}$ in our calculation was considered as a free parameter, setting the neutron fluxes of Maxwell and Fermi spectra at a common energy limit $E_{lim}$ equal:

\begin{equation}
\Phi_{Maxwell} \left( E_{lim} \right) = \Phi_{Fermi} \left( E_{lim} \right).
\label{eq77}
\end{equation}

The high energy neutron spectrum part ($E > E_F$) was defined by fission spectrum~\cite{ref64,ref65,ref66} in our calculations. Therefore for the total volume neutron generation rate $S$ in the expression for the Fermi spectrum~(\ref{eq76}) the following expression may be written:

\begin{equation}
S \left( \vec{r},T,t \right) = \int \limits_{E_F}^{E_n^{max}} \tilde{P} \left( \vec{r},E,T,t \right) \left[ \sum \limits_i \nu_i (E) \cdot \Phi \left( \vec{r},E,T,t \right) \cdot \sigma_f^i (E,T) \cdot N_i \left(\vec{r},T,t \right) \right] dE,
\label{eq78}
\end{equation}

\noindent where $E_n^{max}$ is the maximum energy of neutron fission spectrum (usually taken as $E_n^{max} \approx 10~MeV$), $E_F$ is the neutron energy, below which the moderating neutrons spectrum is described as Fermi spectrum (usually taken as $E_F \approx 0.2~MeV$); $\tilde{P} \left( \vec{r},E,T,t \right)$ is the probability for the neutron not to leave the boundaries of the fissile medium, which depends on the fissile material geometry and conditions at its border as well (e.g. presence of a reflector).

\begin{figure}
  \begin{center}
    \begin{minipage}[c]{0.45\linewidth}
      \begin{center}
      \includegraphics[width=7cm]{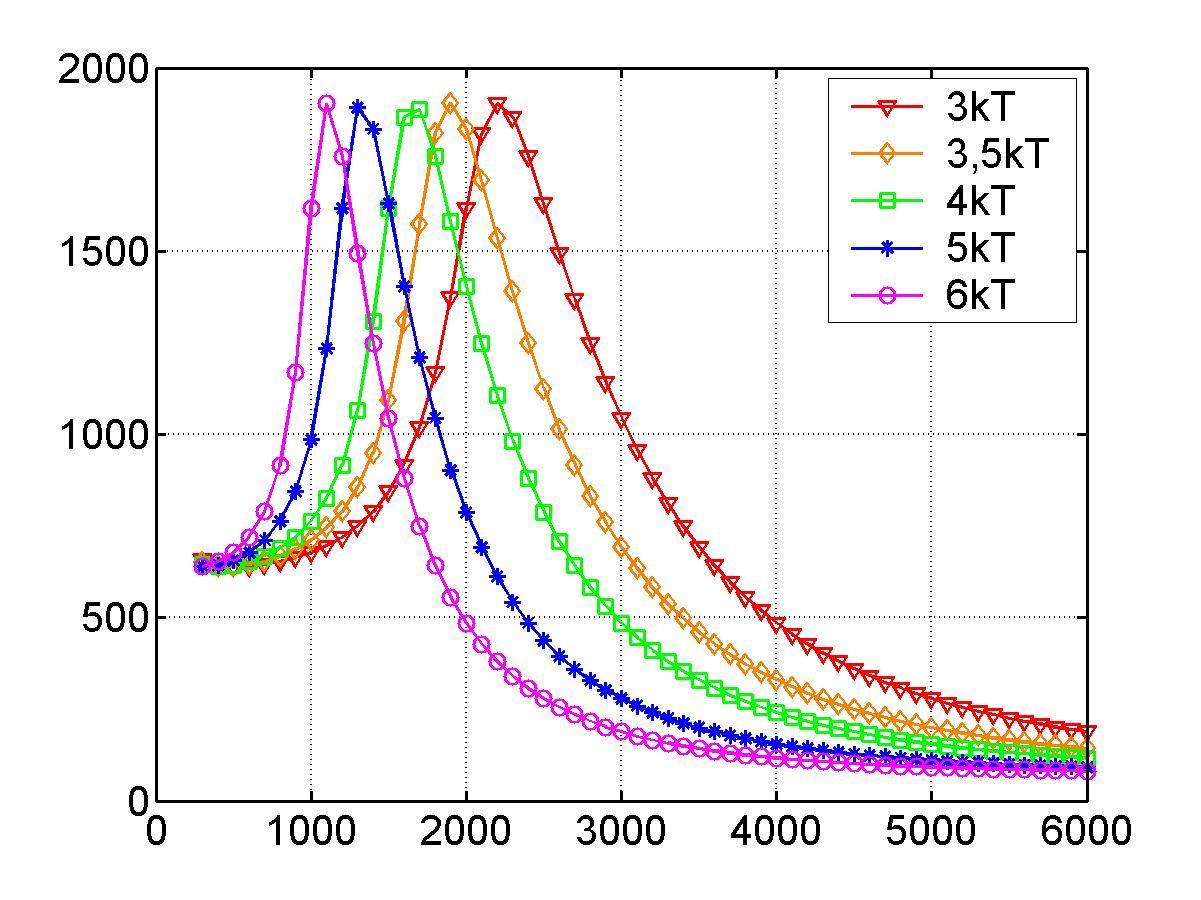} \\ a)
      \end{center}
    \end{minipage}
    \hfill
    \begin{minipage}[c]{0.45\linewidth}
      \begin{center}
      \includegraphics[width=7cm]{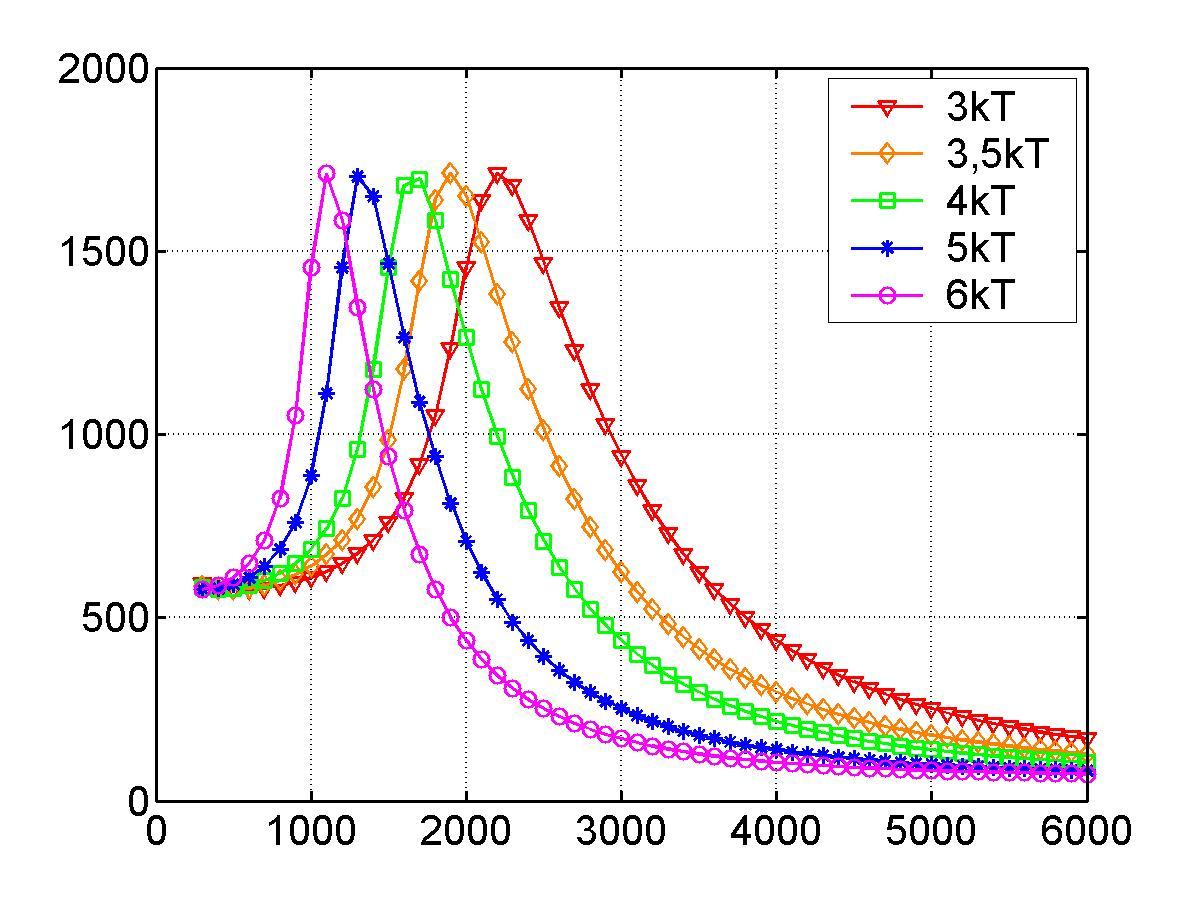} \\ b)
      \end{center}
    \end{minipage}
    \caption{Temperature dependences for the fission cross-section (a) and radioactive capture cross-section (b) for $^{239}Pu$, averaged over the Maxwell spectrum, on the Maxwell and Fermi spectra joining energy and $\eta = 1.8$ (see (\ref{eq75})).}
    \label{fig12}
  \end{center}
\end{figure}

\begin{figure}
  \begin{center}
        \begin{minipage}[c]{0.45\linewidth}
      \begin{center}
      \includegraphics[width=7cm]{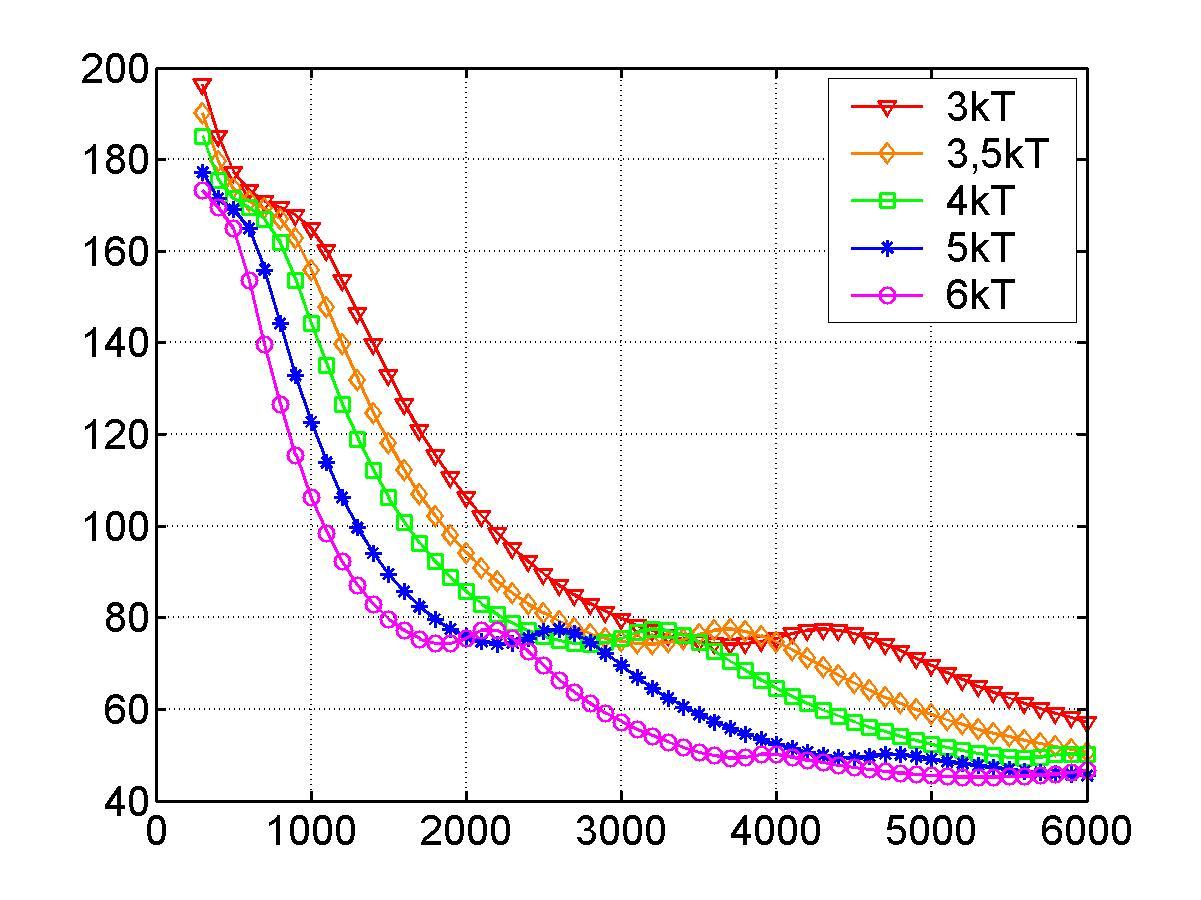} \\ a)
      \end{center}
    \end{minipage}
    \hfill
    \begin{minipage}[c]{0.45\linewidth}
      \begin{center}
      \includegraphics[width=7cm]{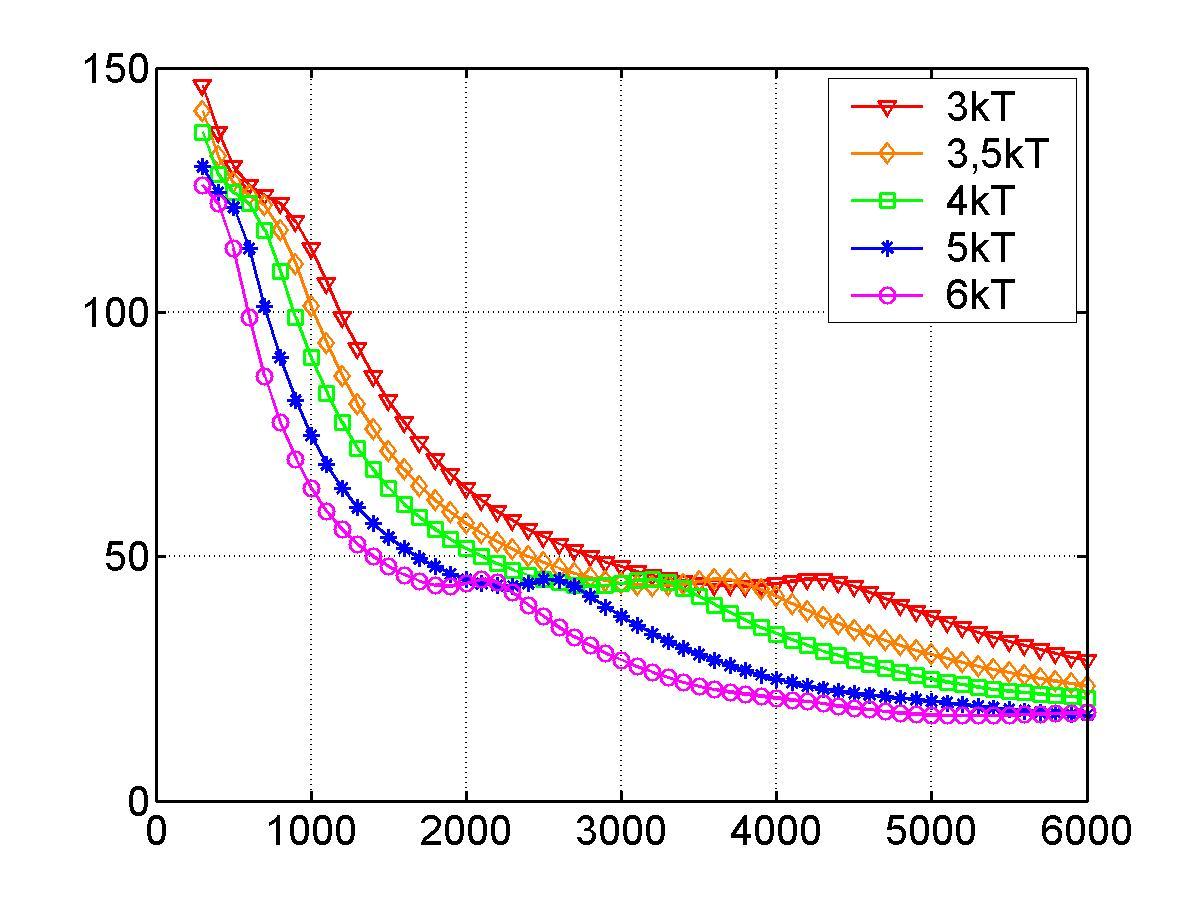} \\ b)
      \end{center}
    \end{minipage}
    \caption{Temperature dependences for the fission cross-section (a) and radioactive capture cross-section (b) for $^{235}U$, averaged over the  Maxwell spectrum, on the Maxwell and Fermi spectra joining  energy and $\eta = 1.8$ (see (\ref{eq75})).}
    \label{fig13}
  \end{center}
\end{figure}

\begin{figure}
  \begin{center}
        \begin{minipage}[c]{0.45\linewidth}
      \begin{center}
      \includegraphics[width=7cm]{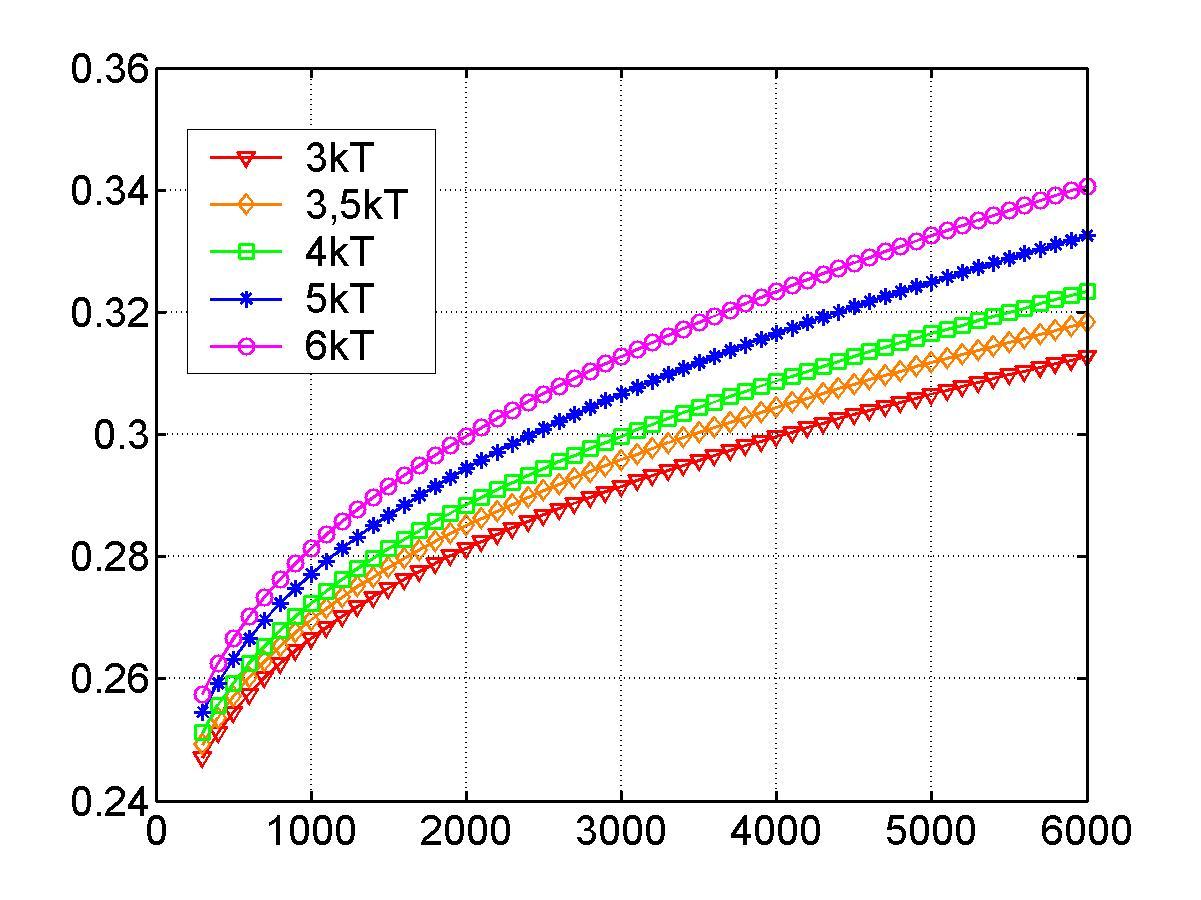} \\ a)
      \end{center}
    \end{minipage}
    \hfill
    \begin{minipage}[c]{0.45\linewidth}
      \begin{center}
      \includegraphics[width=7cm]{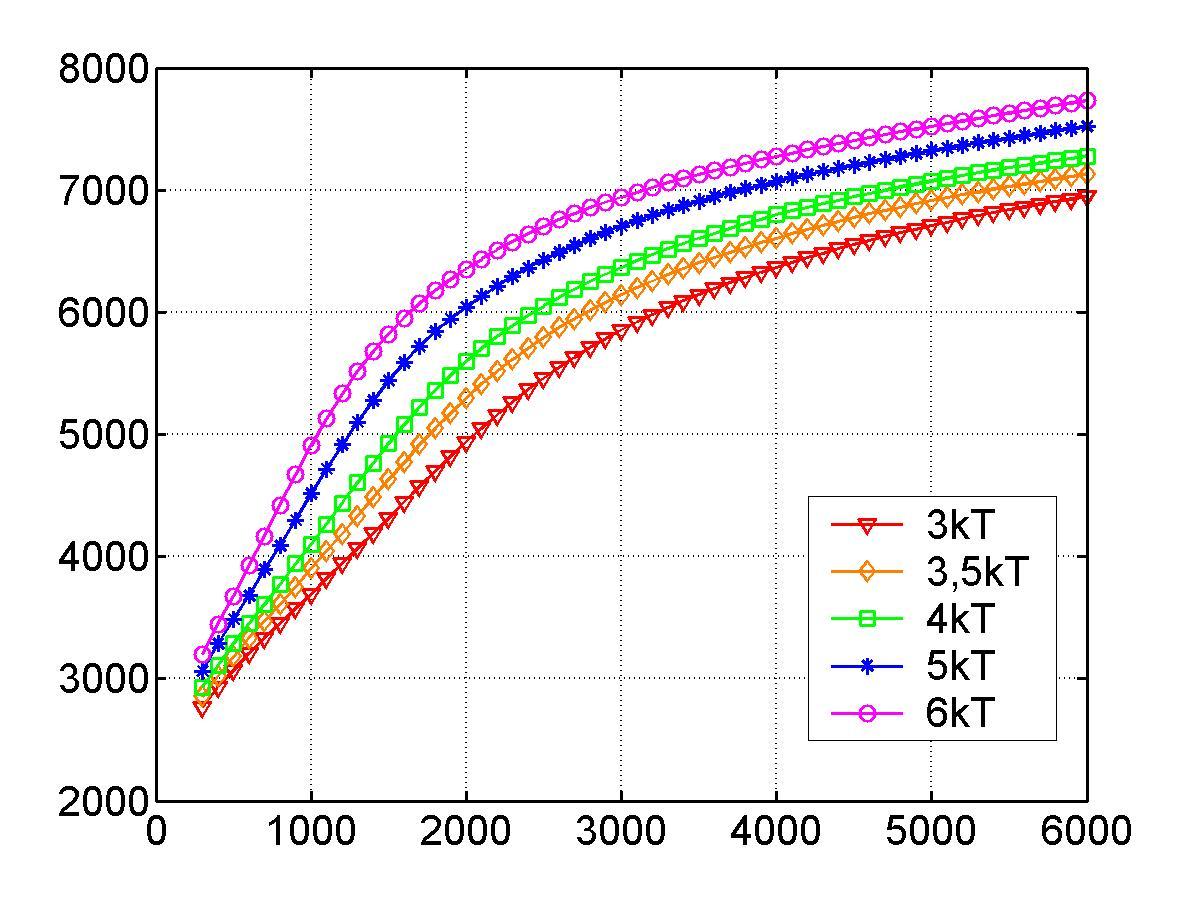} \\ b)
      \end{center}
    \end{minipage}
    \caption{Temperature dependences for the fission cross-section (a) and radioactive capture cross-section (b) for $^{238}_{92}$U, averaged over the  combined Maxwell and Fermi spectra depending on Maxwell and Fermi spectra joining  energy and $\eta = 1.8$ (see (\ref{eq75})).}
    \label{fig14}
  \end{center}
\end{figure}

The obtained calculation results show that the cross-sections averaged over the spectrum may increase (Fig.~\ref{fig12} for $^{239}Pu$ and Fig.~\ref{fig14} for $^{238}U$) as well as decrease (Fig.~\ref{fig13} for $^{235}U$) with fissile medium temperature increase. As follows from the obtained results, the arbitrariness in selection of the limit energy for joining Maxwell and Fermi spectra does not alter the character of these dependences evolution significantly.

This can be justified by the fact that $^{239}$Pu resonance region starts from significantly lower energies than that of $^{235}$U and with fuel temperature increase, the neutron gas temperature increases producing Maxwell's neutron distribution maximum shift to higher neutron energies. In other words, the neutron gas spectrum hardening, when more neutrons fit into resonance area of $^{239}$Pu, is the cause of the averaged cross-sections growth.

For $^{235}$U this process in not as significant because its resonance region is located at higher energies. As a result, the $^{235}$U  neutron gas spectrum hardening related to fuel temperature increase (in the considered interval) does not result in a significant increase of the number of neutrons fitting into the resonance region. Therefore according to the known dispersion relations for $^{235}$U giving the neutron reactions cross-sections behaviour depending on their energy $E_n$ for non-resonance areas, we observe a dependence for the averaged cross-sections $\sigma_{nb} \sim 1/\sqrt{E_n}$.

The data on the averaged fission and capture cross-sections  of $^{238}$U presented at Fig.~\ref{fig14} show that the averaged fission cross-section for $^{238}$U is almost insensitive to the neutron spectrum hardening caused by the fuel temperature increase, because of the high fission threshold $\sim$1~MeV (see fig.~\ref{fig14}a). On the other hand, they confirm the dependence of the capture cross-section on temperature, because its resonance region is located as low as for $^{239}$Pu. Obviously, fuel enrichment with  $^{235}$U essentially makes no difference in this case, because the averaged cross-sections for $^{235}$U, as described above, behave in a standard way.

\begin{figure}
  \begin{center}
    \includegraphics[width=10cm]{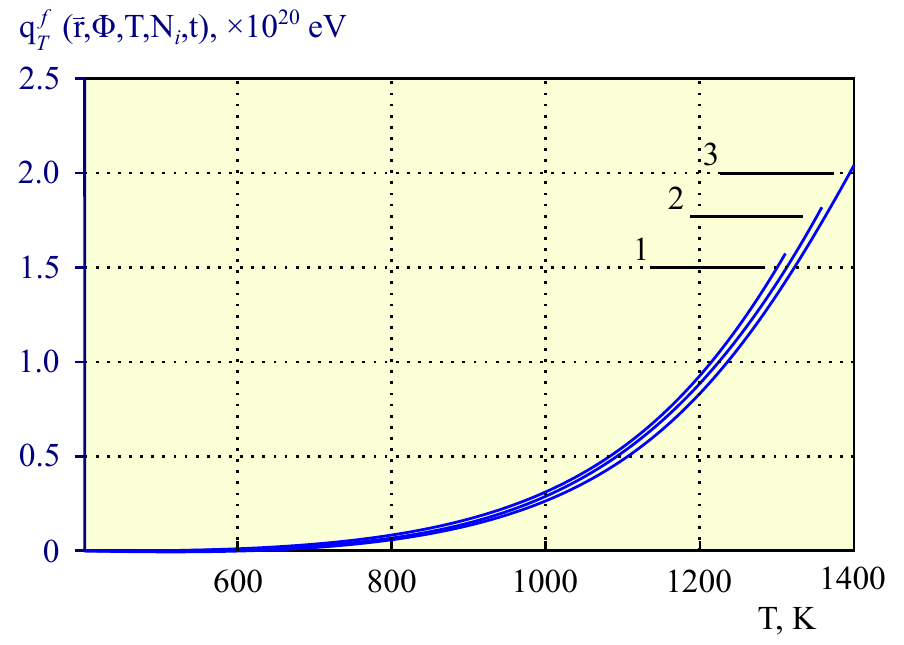}
    \caption{Dependence of the heat source density  $q_T ^f \left( r, \Phi, T, N_i, t \right)~[eV]$ on the fissile medium temperature (300-1400K) for  several compositions of uranium-plutonium medium (1 - 10\%~Pu; 2 - 5\%~Pu; 3 - 1\%~Pu) at a constant neutron flux density $\Phi = 10^{13}~n/(cm^2 \cdot s)$.}
    \label{fig15}
  \end{center}
\end{figure}

And finally we performed a computer estimate of the heat source density dependence $q_T^f \left(\vec{r},\Phi,T,t \right)$ (\ref{eq74}) on temperature for different compositions of uranium-plutonium fissile medium with a constant neutron flux density, presented at fig.~\ref{fig15}. We used the dependences presented above at Fig.\ref{fig12}-\ref{fig14} for these calculations. Let us note that our preliminary calculations were made not taking into account the change in composition and density of the fissile uranium-plutonium medium that is a direct consequence of the constant neutron flux assumption.

The necessity of such assumption is caused by the following. It is obvious that for a reasonable description of the neutron source density $q_T^f \left(\vec{r},\Phi,T,t \right)$ (\ref{eq74}) dependence on temperature, a system of three equations must be solved. Two of them correspond to the neutron kinetics equation (flux and fluence) and to the system of equations for the parental and child nuclides nuclear density kinetics (e.g. see~\cite{ref29,ref26}), while the third equation corresponds to a heat transfer equation of~(\ref{eq69}) type. However, some serious difficulties arise at this point, associated with the limited computational capabilities available. And here is why.

One of the principal physical peculiarities of TWR is the fact~\cite{ref20} that fluctuation residuals of plutonium (or $^{233}$U in Th-U cycle) over its critical concentration burn out for the time comparable with the reactor lifetime of a neutron $\tau_n (x,t)$ (not considering delayed neutrons), or at least comparable with the reactor period\footnote{The reactor period by definition equals to $T(x,t) = \tau_n (x,t) / \rho(x,t)$, i.e. is a ratio of the reactor neutron lifetime to reactivity.} $T(x,t)$ (considering delayed neutrons). Meanwhile, the new plutonium (or $^{233}$U in Th-U cycle) is formed in a few days (or a month) and not at once. It means~\cite{ref20} that numerical calculation must be performed with a temporal step around 10$^{-6}$-10$^{-7}$ for the case of not taking into account the delayed neutrons and $\sim$10$^{-1}$-10$^{0}$ otherwise. At first glance, taking into account the delayed neutrons, according to~\cite{ref20}, really "saves the day", however it is not always true. If the heat transfer equation contains a significantly non-linear source, then in the case of a blow-up mode, the temperature at some conditions may grow extremely fast and in 10-20 steps (with time step 10$^{-6}$-10$^{-7}$~s) reaches the critical amplitude that may lead to (as a minimum) a solution stability loss or (as a maximum) blow-up bifurcation of the phase state, almost unnoticeable with a rough time step.

Here we should also mention that modern scientific literature has absolutely no theoretical or experimental data on heat capacity $c_p$ and heat conductivity $\aleph$ of the fissile material for the temperatures over 1500K, which makes the further model calculations quite problematic.

According to these remarks, and considering the goal and format of this paper, we didn't aim at finding the exact solution of some specific system of three joined equations described above. However, we found it important to illustrate -- at the qualitative level -- the consequences of the possible blow-up modes in case of non-linear heat source existence in the heat transfer equation. As it was described above, we performed the estimate computer calculations of the heat source density $q_T^f \left(\vec{r},\Phi,T,t \right)$~(\ref{eq74}) dependence on temperature in 300-1400K range for selected compositions of uranium-plutonium fissile medium at a constant neutron flux (Fig.~\ref{fig15}).

The obtained dependences for the heat source density $q_T^f \left(\vec{r},\Phi,T,t \right)$ were successfully approximated by a power function of temperature with an exponent of 4 (Fig.~\ref{fig15}). In other words, we obtained a heat transfer equation with a significantly nonlinear heat source in the following form:

\begin{equation}
q_T (T) = const \cdot T^{(1 + \delta)},
\label{eq79}
\end{equation}

\noindent where $\delta > 1$ in the case of a non-linear heat conductivity dependence on temperature \cite{ref55,ref56,ref57,ref58,ref59}. The latter means that the solutions of the heat transfer equation~(\ref{eq69}) describe the so-called Kurdyumov blow-up modes \cite{ref55,ref56,ref57,ref58,ref59,ref60}, i.e. such dynamic modes when one of the modeled values (e.g. temperature) turns into infinity for a finite time interval. As noted before, in reality instead of reaching infinite values, a corresponding phase transition is observed (a final phase of the parabolic temperature growth), that requires a separate model and is a basis for an entirely new problem.

Mathematical modeling of blow-up  modes was performed mainly using Mathematica~5.2-6.0, Maple~10, Matlab~7.0, utilizing multiprocessor calculations for effective application. Runge-Kutta method of 8-9$^{th}$ order and the numerical methods of lines~\cite{ref67} were also applied to the calculations. The numerical error estimate did not exceed 0.01\%. The coordinate and temporal step were variable and chosen by the program in order to fit the given error at every calculation step.

\begin{figure}
  \begin{center}
    \includegraphics[width=10cm]{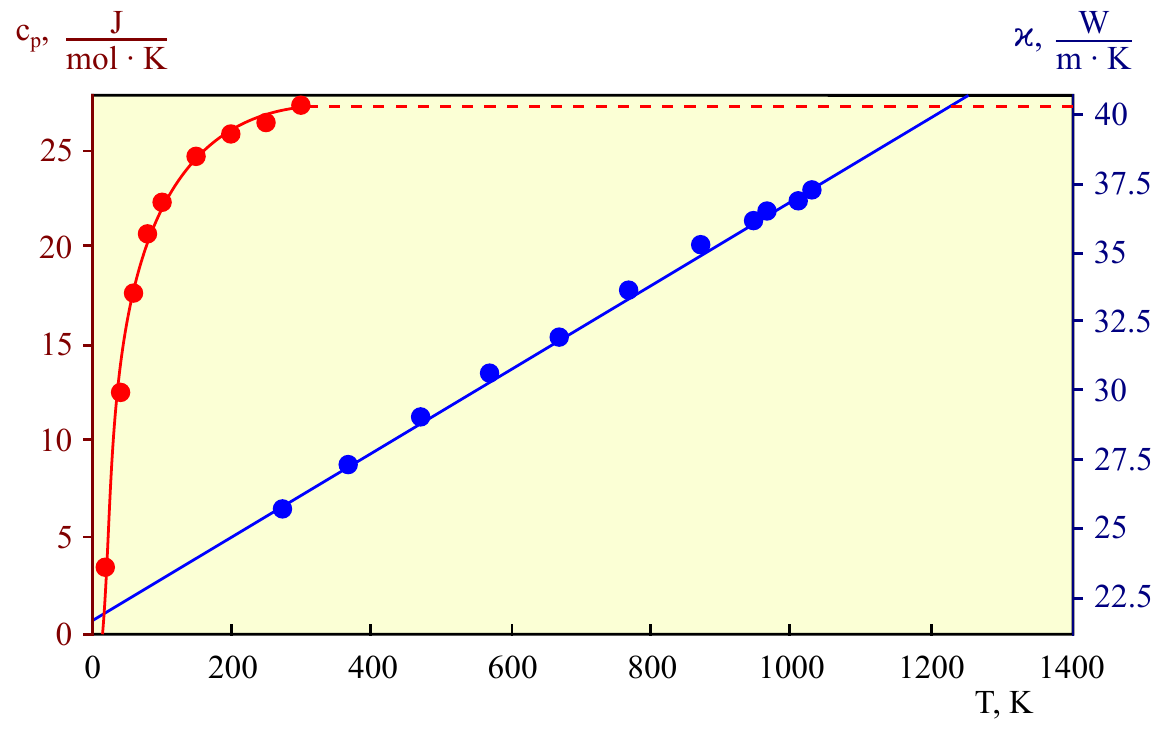}
    \caption{Temperature dependence of the  fissile material heat capacity $c_P$ and thermal conductivity $\chi$. Points represent the experimental values for the heat capacity and thermal conductivity of $^{238}$U.}
    \label{fig16}
  \end{center}
\end{figure}

Below we give the solutions for the heat transfer equation~(\ref{eq69}) with nonlinear exponential heat source~(\ref{eq79}) in  uranium-plutonium fissile medium for the boundary and initial parameters corresponding to those of the technical reactors. The calculations were done for a cube of fissile material with different spatial size, boundary and initial temperature values. Since the temperature dependences of the heat source densities were obtained without account for changing composition and density of the uranium-plutonium fissile medium, different blow-up modes can take place (HS-mode, S-mode, LS-mode) depending on the ratio between the exponents of the temperature dependences of thermal conductivity and heat source according to \cite{ref55,ref56,ref57,ref58,ref59,ref60}. Therefore we considered cases for 1$^{\text{st}}$, 2$^{\text{nd}}$ and 4$^{\text{th}}$ temperature order sources. Here the power of the source also varied by varying the proportionality factor in~(\ref{eq79}) ($const = 1.00 J / (cm^3 \cdot s \cdot K$) for 1$^{\text{st}}$ temperature order source; $0.10 J / (cm^3 \cdot s \cdot K^2)$, $0.15~J / (cm^3 \cdot s \cdot K^2)$ and $1.00~J / (cm^3 \cdot s \cdot K^2)$ for 2$^{\text{nd}}$ temperature order source; $1.00~J / (cm^3 \cdot s \cdot K^4)$ for 4$^{\text{th}}$ temperature order source).
While calculating the the heat capacity $c_p$ (Fig.~\ref{fig16}a) and thermal conductivity $\aleph$ (Fig.~\ref{fig16}b) dependences on the fissile medium temperature in 300-1400K range, the specified parameters were given by the analytic expressions, obtained by approximation of the experimental data for$^{238}$U based on the polynomial progression:

\begin{align}
c_p (T) \approx - 7.206 + 0.64 T & - 0.0047 T^2 + 0.0000126 T^3 + \nonumber \\
& + 2.004 \cdot 10^{-8} T^4 - 1.60 \cdot 10^{-10} T^5 - 2.15 \cdot 10^{-13} T^6,
\label{eq80}
\end{align}

\begin{equation}
\aleph (T) \approx 21.575 + 0.0152661 T.
\label{eq81}
\end{equation}

\begin{figure}
  \begin{center}
    \includegraphics[width=15cm]{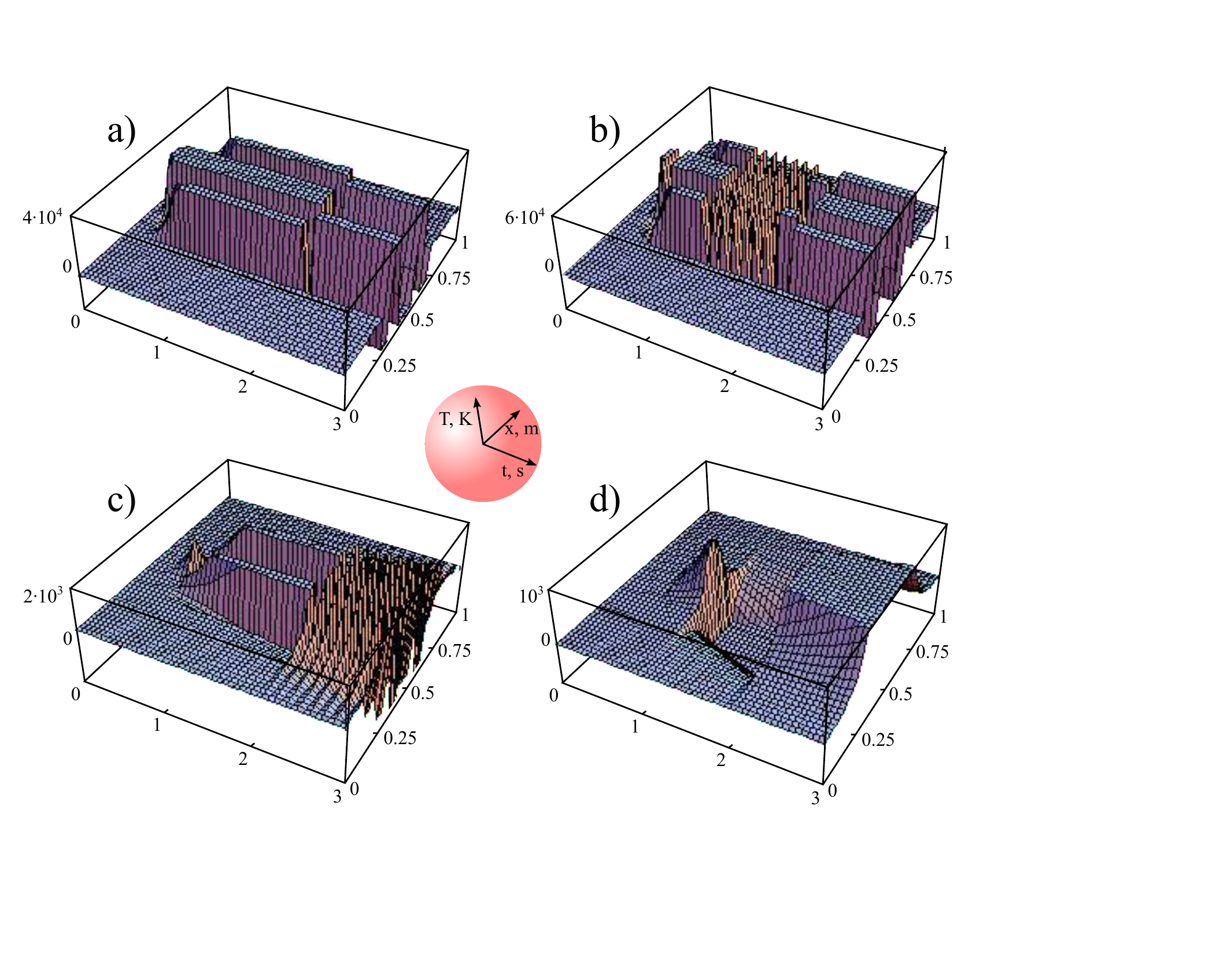}
    \caption{Heat transfer equation~(\ref{eq69}) solution for 3D case (crystal sizes 0.001$\times$0.001$\times$0.001~mm; initial and boundary temperatures equal to 100K): a) The source is proportional to the 4$^{\text{th}}$ order of temperature; $const$ =1.00 $J / (cm^3 \cdot s \cdot K^4)$, heat capacity and thermal conductivity are constant and equal to 11.5 $ J / (K \cdot mol)$ and 27.5 $W / (m \cdot K)$ respectively; b) the source is proportional to the 4$^{\text{th}}$ order of temperature; $const$ =1.00 $J / (cm^3 \cdot s \cdot K^4)$; c) The source is proportional to the 2$^{\text{nd}}$ order of temperature; $const$ =1.00 $J / (cm^3 \cdot s \cdot K^2)$; d) the source is proportional to the 2$^{\text{nd}}$ order of temperature; $const$ = 0.10 $J / (cm^3 \cdot s \cdot K^2)$. Note: in cases b)- d) the heat capacity and thermal conductivity were determined by~(\ref{eq80}) and~(\ref{eq81}) respectively.}
    \label{fig17}
  \end{center}
\end{figure}

And finally a solution of the heat transfer equation~(\ref{eq69}) was obtained for the constant thermal conductivity ($27.5~W / (m \cdot K)$) and heat capacity ($11.5~J / (K \cdot mol)$) values, presented in Fig.~\ref{fig17}a, as well as the solutions of the heat transfer equation considering their temperature dependences (Fig.~\ref{fig17}b-d).

Preliminary results point directly to a possibility of the local melting of the uranium-plutonium fissile medium, having melting temperature almost identical to that of $^{238}$U, that equals 1400K (Fig.~\ref{fig16}a-d). Moreover, these regions of the local melting are not the areas of the so-called thermal spikes~\cite{ref68}, and probably are the anomalous areas of the uranium surface melting observed by Laptev and Ershler~\cite{ref69} that were also mentioned in~\cite{ref70}. More detailed analysis of the probable temperature scenario associated with the blow-up modes is discussed below.

\section{Physical peculiarities of the blow-up regimes in neutron-multiplying media}
\label{sec3}

Earlier we noted the fact that due to the coolant loss at nuclear reactors during Fukushima nuclear accident the fuel was melted, or in other words, temperature inside the active zone at some moment reached the melting temperature of the uranium-oxide fuel, i.e. $\sim$3000$^{\circ}$C. 

On the other hand, we already know that the coolant loss may become a cause of the nonlinear heat source formation inside the nuclear fuel and thus become a cause of the temperature and neutron flux blow-up mode occurrence. A natural question arises as to whether it is possible to use such blow-up mode (in terms of temperature and neutron flux) for the initiation of certain controlled physical conditions, under which  the nuclear burning wave would regularly "experience" the so-called controlled blow-up regime. It is quite difficult to answer this question definitely, because such fast process has some physical vaguenesses, any of which can become experimentally unsurmountable for the its controlling.

Nevertheless such process is very elegant and beautiful from the physical point of view and therefore requires more detailed phenomenological description. Let us try to make it in short.

As one can see from the plots of the capture and fission cross-sections evolution for $^{239}$Pu (Fig.~\ref{fig12}), the blow-up mode may develop actively at $\sim$1000-2000K (depending on the real value of Fermi and Maxwell spectra joining boundary), but it returns to almost the initial cross-sections values at temperatures over 2500-3000K. If we turn on some effective heat sink at that point, the fuel may return to its initial temperature state. However, while the blow-up mode develops, the fast neutrons already penetrate to the adjacent fuel areas, where the new fissile material starts accumulating and so on (see cycles~(\ref{eq69}) and~(\ref{eq70})). After some time the similar blow-up mode will develop in this adjacent area and everything starts over again. In other words, such hysteresis blow-up mode, closely time-conjugated to a heat takeoff procedure, will appear against the background of a stationary nuclear burning wave in a form of the periodic impulse bursts.

In order to demonstrate the marvelous power of such process, we investigated the heat transfer equation with non-linear exponential heat source in uranium-plutonium fissile medium with the boundary and initial parameters emulating the heat takeoff process. In other words, we investigated the blow-up modes in the fast Feoktistov-type uranium-plutonium reactor~\cite{ref3} where the temperature was deliberately fixed at 6000K inside and outside the boundary.

This temperature is defined by the following important question: "Is it possible to obtain a solution, i.e. the spacio-temporal temperature distribution not in a form of a $\delta$-function at some local spatial area, but as some kind of a stationary and limited by amplitude solitary wave under such conditions (6000K), which emulate the time-conjugated heat takeoff (see Fig.~\ref{fig12})?" As shown below, such suggestion proved productive.

Below we present some calculation characteristics and parameters. During these calculations we used the following expression for the heat conductivity coefficient:

\begin{equation}
\aleph = 0.18 \cdot 10^{-4} \cdot T  \nonumber,
\end{equation}

\noindent which was obtained using Wiedemann–Franz law and the data on electric conductivity of metals at temperature 6000K~\cite{ref71}. Specific heat capacity at constant pressure was set to $c_p \approx 6~cal/(mol \cdot deg)$ according to Dulong and Petit law.

\begin{figure}
  \begin{center}
    \includegraphics[width=15cm]{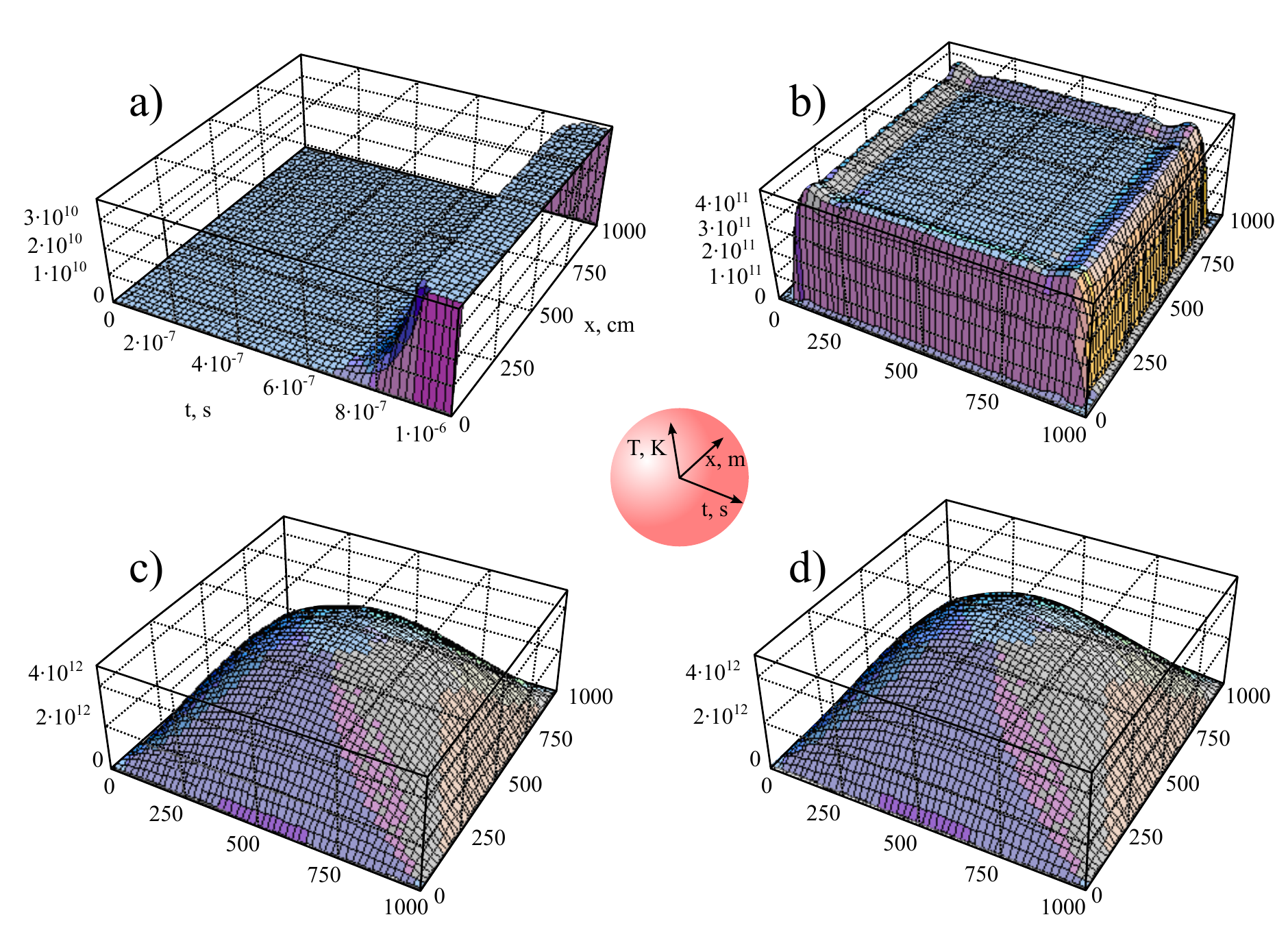}
    \caption{Heat transfer equation solution for a model reactor (source $\sim$2$^{nd}$ order temperature dependence, $const$ = 4.19 $J / (cm^3 \cdot s \cdot K^2)$; Initial and boundary temperatures equal to 6000K;  fissile medium is a cube 10$\times$10$\times$10~m. The presented results correspond to the following times of the temperature field evolution: (a) (1-10)$\cdot 10^{-7}$~s, (b) $10^{-6}$~s, (c) 0.5~s, (d) 50~s.}
    \label{fig18}
  \end{center}
\end{figure}

The fissile uranium-plutonium medium was modeled as a cube of the size 10.0$\times$10.0$\times$10.0~m (Fig.\ref{fig18}) during the calculations. Here we used the 2nd order temperature dependence for the heat source (see~(\ref{eq79})).
 
And finally Fig.~\ref{fig18}a-d present a set of solutions of the heat transfer equation~(\ref{eq69}) with nonlinear exponential heat source (\ref{eq79}) in uranium-plutonium fissile medium with boundary and initial conditions emulating such process of heat takeoff that initial and boundary temperatures remain constant and equal to 6000K.

It is important to note here, that the solution set presented at Fig.~\ref{fig18} demonstrates  the temporal evolution of the solution to its "stationary" state quite clearly. This is achieved using the so-called "magnifying glass" approach when the solutions of the same problem are deliberately investigated at the different timescales. For example, Fig.~\ref{fig18}a shows the solution at the time scale $t \in [0,10^{-6}~s]$, while Fig.~\ref{fig18}b describes the spatial solution of the problem (temperature field) for $t = 10^{-6}~s$. The Fig.~\ref{fig18}c-d present the solution (the spatial temperature distribution) at $t = 0.5~s$ and $t = 50~s$. 

As one can see, the solution (Fig.~\ref{fig18}d) is completely identical to the previous (Fig.\ref{fig18}c), i.e. to the distribution established in the medium in 0.5 seconds, which allows us to make a conclusion about the temperature field stability, starting from some moment. It is interesting that the established temperature field creates the conditions enough for thermonuclear synthesis reaction, i.e. reaching 10$^{8}$K, and the time of such temperature field existence is not less than 50~s. These conditions are highly favorable for a stable thermonuclear burning given the necessary nuclei concentration to enter the thermonuclear synthesis reaction.

That said, one should remember that the results of the current chapter are only demonstrative, because their accuracy is very relative and requires careful investigations involving the necessary computational resources. However, qualitative peculiarities of these solutions should attract the researchers' attention to the nontrivial properties of the blow-up modes, at least, with respect to the obvious problem of internal TWR safety violation.

\section{Conclusion}
\label{sec4}

Below we give short conclusions stimulated by significant problems, that can be formulated in the following form.

\begin{enumerate}
\item \textbf{The consequences of the anomalous $^{238}$U and $^{239}$Pu cross-sections behavior with temperature increase.} It is shown that the capture and fission cross-sections of $^{238}$U and $^{239}$Pu manifest a monotonous growth in 1000-3000K range. Obviously, such anomalous temperature dependence of $^{238}$U and $^{239}$Pu cross-sections changes the neutron and heat kinetics drastically in the nuclear reactors, and in TWR in particular. It becomes crucial to know their influence on kinetics of heat transfer because it may become the cause of a positive feedback with neutron kinetics, which may lead not only to undesirable loss of solution stability (the nuclear burning wave), but also to a trivial reactor runaway with a subsequent nontrivial disaster.

\item \textbf{Blow-up modes and the problem of nuclear burning wave stability.} One of the causes of a possible fuel temperature growth may lie, for instance, in a deliberate or spontaneous coolant loss, analogous to what happened during the Fukushima nuclear accident. As shown above, the coolant loss may become a cause of the nonlinear heat source formation in the nuclear fuel, and consequently emergence of the mode with the temperature and neutron flux blow-up. In our opinion, the preliminary investigation of the heat transfer equation with nonlinear heat source points to an extremely important phenomenon of the anomalous of the temperature and neutron flux blow-up modes behavior. This result poses a natural nontrivial problem of fundamental nuclear burning wave stability and,  orrespondingly, of a physically reasonable application of Lyapunov method to this problem, which is a basis for the \textit{motion stability} theory, and thus a reliable basis for justification of the Lyapunov functional minimum existence.

It is shown that some variants of solution stability loss are caused by anomalous nuclear fuel temperature evolution. They can be not only the cause of  TWR internal safety loss, but can lead to a new stable mode when nuclear burning wave would periodically "experience" the so-called controlled blow-up regime through a bifurcation of states (which is very important!). At the same time, it is noted that such fast (blow-up regime) process has a number of physical  vaguenesses, any of which may happen to be experimentally insurmountable for the control of such process.

\item \textbf{On-line remote neutrino diagnostics of intra-reactor processes.} Due to the fact that a high-power TWR or a nuclear fuel transmutation reactor are the projects of the single-load and fuel burn-up with a subsequent burial of the reactor apparatus, there is an obvious necessity for remote system of neutrino monitoring of the neutron-nuclear burning wave in normal operation mode and control of the neutron kinetics in emergency situation. The calculation of the spatio-temporal distribution of the isotope composition in the TWR active zone in the framework of the inverse problem of neutrino diagnostics of intra-reactor processes is presented in~\cite{ref73,ref17,ref75} in detail.

\end{enumerate}

\bibliographystyle{unsrtnat}
\bibliography{TravelingWaveReactor}

\end{document}